\documentclass[preprint]{iucr}              

     \journalcode{}              

\RequirePackage{graphicx}
\usepackage{verbatim,xcolor,soul}
\usepackage{amsmath,lineno,amssymb,amsthm,mathtools,url}

\theoremstyle{definition}
\newtheorem{dfn}{Definition}[section]
\newtheorem{pro}[dfn]{Problem}
\newtheorem{thm}[dfn]{Theorem}

\newtheorem{lem}[dfn]{Lemma}
\newtheorem{exa}[dfn]{Example}

\newcommand{\R}{\mathbb R}
\newcommand{\Z}{\mathbb Z}

\newcommand{\al}{\alpha}
\newcommand{\be}{\beta}

\newcommand{\de}{\delta}

\newcommand{\La}{\Lambda}
\newcommand{\si}{\sigma}
\newcommand{\ep}{\varepsilon}

\newcommand{\SO}{\mathrm{SO}}
\newcommand{\Or}{\mathrm{O}}

\newcommand{\VF}{\mathrm{VF}}
\newcommand{\CF}{\mathrm{CF}}

\newcommand{\RF}{\mathrm{RF}}

\newcommand{\Oct}{\mathrm{Oct}}
\newcommand{\FT}{\mathrm{FT}}
\newcommand{\ST}{\mathrm{FT}}

\newcommand{\QT}{\mathrm{QT}}
\newcommand{\CM}{\mathrm{RM}}
\newcommand{\RM}{\mathrm{RM}}
\newcommand{\RFL}{\mathrm{RFL}}
\newcommand{\OSI}{\mathrm{OSI}}
\newcommand{\LISP}{\mathrm{LISP}}

\newcommand{\bs}{\hfill $\blacksquare$}
\newcommand{\lra}{\leftrightarrow}

\newcommand{\vl}{\,:\,}
\newcommand{\mat}[6]{\left(\begin{array}{ccc}
#1 & #2 & #3 \\ #4 & #5 & #6 \end{array}\right)}

\begin{document}                  



\title{A complete and continuous map of the Lattice Isometry Space for all 3-dimensional lattices}
\shorttitle{A complete and continuous map of the Lattice Isometry Space}


\author[a]{Matthew}{Bright}{}{}
\author[a]{Andrew I}{Cooper}{}{}
\cauthor[a]{Vitaliy}{Kurlin}{vitaliy.kurlin@gmail.com}{}

\aff[a]{Materials Innovation Factory, University of Liverpool, \country{UK}}









\maketitle                        

\begin{synopsis}

\end{synopsis}

\begin{abstract}
This paper extends the recently obtained complete and continuous map of the Lattice Isometry Space (LISP) to the practical case of dimension 3.
A periodic 3-dimensional lattice is an infinite set of all integer linear combinations of basis vectors in Euclidean 3-space. 
Motivated by crystal structures determined in a rigid form, we study lattices up to rigid motion or isometry, which is a composition of translations, rotations and reflections.
The resulting space LISP consists of infinitely many isometry classes of lattices.
In dimension 3, we parameterise this continuous space LISP by six coordinates and introduce new metrics satisfying the metric axioms and continuity under all perturbations.
This parameterisation helps to visualise hundreds of thousands of real crystal lattices from the Cambridge Structural Database for the first time.
\end{abstract}


\section{Motivations, metric problem and overview of past and new results}
\label{sec:intro}

This paper continues the related work \cite{bright2021easily}, which provided practical motivations for a metric map of the Lattice Isometry Space and then focused on 2-dimensional lattices.
Briefly, since crystal structures are determined in a rigid form, their most fundamental equivalence is rigid motion (any composition of translations and rotations in $\R^3$).
The concept of an isometry (any map preserving Euclidean distances) also includes mirror reflections.
It is a bit more convenient to study the isometry equivalence.
We can easily detect if an isometry preserves an orientation.
\medskip
  
\begin{figure}
\label{fig:lattice_classification}
\caption{
The LISP is bijectively and bi-continuously mapped to root forms of lattices, which are triples of root products between vectors of an obtuse superbase in $\R^2$.}
\includegraphics[width=1.0\textwidth]{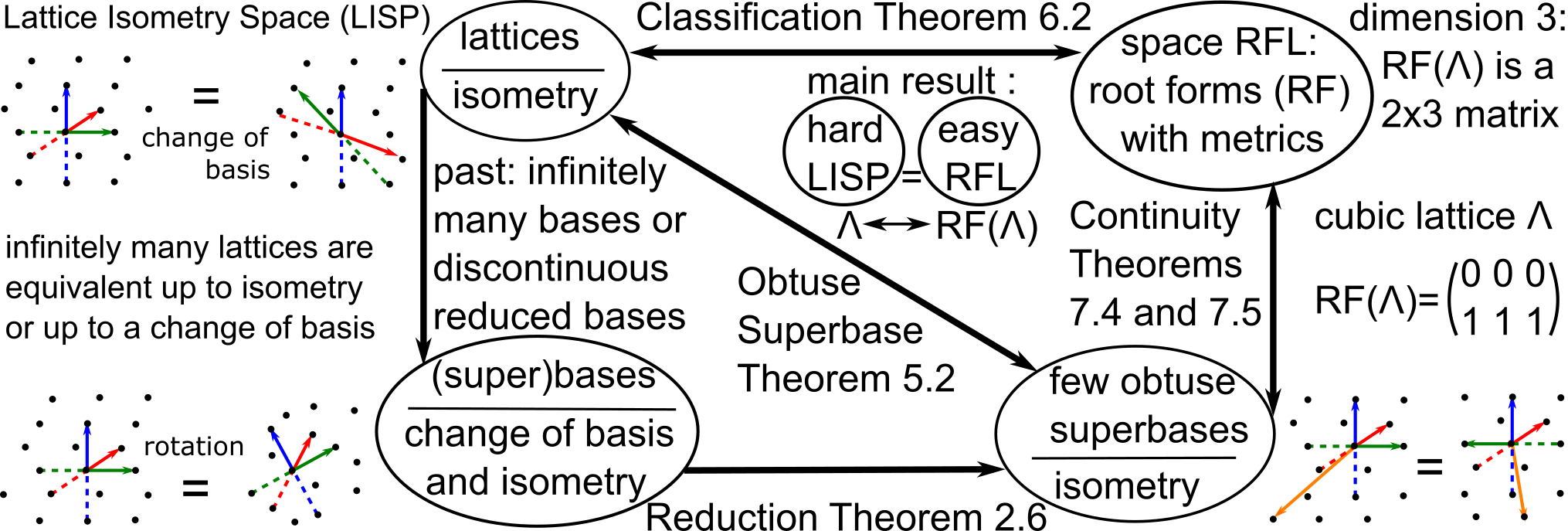}
\end{figure}

Isometry is the fundamental equivalence of lattices due to rigidity of most crystals.
The resulting Lattice Isometry Space (LISP) consists of infinitely many classes, where every class includes all lattices isometric to each other.
Then any transition between lattices is a continuous path in the LISP.
The two square lattices in the top left corner of Fig.~\ref{fig:lattice_classification} have different bases (related by a rotation) but belong to the same isometry class of unit square lattices.
A past approach to uniquely represent any isometry class was to choose a reduced basis (Niggli's reduced cell).
Any such reduction is discontinuous under perturbations of a basis, see \citeasnoun[Theorem~15]{widdowson2022average}.

Metric Problem~\ref{pro:metric} is stated below for any dimension $n\geq 2$.
The main contribution is the extension of the solution for $n=2$ from \cite{bright2021easily} to $n=3$.

\begin{pro}[metric on lattices]
\label{pro:metric}
Find a metric $d(\La,\La')$ on lattices in $\R^n$ such that
\smallskip

\noindent
(\ref{pro:metric}a) 
$d(\La,\La')$ is independent of given primitive bases of lattices $\La,\La'$;
\smallskip

\noindent
(\ref{pro:metric}b) 
the function $d(\La,\La')$ is preserved under any isometry or rigid motion of $\R^n$;
\smallskip

\noindent
(\ref{pro:metric}c)
$d$ satisfies the metric axioms: $d(\La,\La')=0$ if and only if $\La,\La'$ are isometric, symmetry $d(\La,\La')=d(\La',\La)$ and triangle inequality $d(\La,\La')+d(\La',\La'')\geq d(\La,\La'')$;
\smallskip

\noindent
(\ref{pro:metric}d)
$d(\La,\La')$ continuously changes under perturbations of primitive bases of $\La,\La'$;
\smallskip

\noindent
(\ref{pro:metric}e)
$d(\La,\La')$ is computed from reduced bases of $\La,\La'$ in a constant time.
\bs 
\end{pro}

\citeasnoun[section~2]{bright2021easily} has reviewed many past attempts to solve Problem~1.1, especially based on Niggli's reduced cell \cite{niggli1928krystallographische}, whose discontinuity \cite{andrews1980perturbation} fails condition (\ref{pro:metric}d).
We should certainly mention the celebrated efforts of Larry Andrews and Herbert Bernstein \cite{andrews1988lattices,andrews2014geometry,mcgill2014geometry,andrews2019selling}
whose latest advance is the $DC^7$ function comparing lattices by the seven ordered distances from the origin to its closest neighbours \cite{andrews2019space}. 
This function $DC^7$ turns out to be a nearly ideal solution to Problem~\ref{pro:metric},  see details in Example~\ref{exa:dc7=0}.
\medskip

Section~\ref{sec:definitions} formally defines key concepts, most importantly Voronoi domains.
Following \cite{conway1992low}, we remind an obtuse superbase consisting of vectors $v_1,v_2,v_3$ and $v_0=-v_1-v_2-v_3$ in $\R^3$ such that all vectors have non-acute angles, equivalently non-positive scalar products $v_i\cdot v_j\leq 0$.
Section~\ref{sec:invariants3d} introduces the root products $r_{ij}=\sqrt{-v_i\cdot v_j}$, coordinates on the space of root forms of lattices (RFL).
\medskip

This space $\RFL$ provides a complete and continuous parameterisation of the Lattice Isometry Space (LISP) as follows.
Theorem~\ref{thm:superbases/isometry} substantially reduces the ambiguity of lattice representations by infinitely many bases to only very few obtuse superbases, see the bottom right corner in Fig.~\ref{fig:lattice_classification}.
Theorem~\ref{thm:classification3d} proves completeness of root forms by establishing an invertible 1-1 map $\LISP\lra\RFL$.
Theorems~\ref{thm:superbases->root_forms} and~\ref{thm:root_forms->superbases} prove that this 1-1 map is continuous in both directions.
As a result, we have a complete and continuous metric map on the isometry space of lattices ($\LISP$) in $\R^3$.

\section{Basic definitions and Conway-Sloane's results for lattices}
\label{sec:definitions}

Any point $p$ in Euclidean space $\R^n$ can be represented by the vector from the origin $0\in\R^n$ to $p$.
So $p$ may also denote this vector, though an equal vector $p$ can be drawn at any initial point.
The \emph{Euclidean} distance between points $p,q\in\R^n$ is $|p-q|$.

\begin{dfn}[a lattice $\La$, a unit cell $U$]
\label{dfn:periodic_set}
Let vectors $v_1,\dots,v_n$ form a linear {\em basis} in $\R^n$ so that if $\sum\limits_{i=1}^n c_i v_i=0$ for some real $c_i$, then all $c_i=0$.
Then a {\em lattice} $\La$ in $\R^n$ consists of all linear combinations $\sum\limits_{i=1}^n c_i v_i$  with integer coefficients $c_i\in\Z$.
The parallelepiped $U(v_1,\dots,v_n)=\left\{ \sum\limits_{i=1}^n c_i v_i \vl c_i\in[0,1) \right\}$ is a \emph{primitive unit cell} of $\La$.
\bs
\end{dfn}

The (signed) volume $V$ of a unit cell $U(v_1,\dots,v_n)$ equals the determinant of the $n\times n$ matrix with columns $v_1,\dots,v_n$.
The sign of $V$ is used to define an \emph{orientation}.

\begin{dfn}[isometry, orientation and rigid motion]
\label{dfn:isometry}
An \emph{isometry} is any map $f:\R^n\to\R^n$ such that $|f(p)-f(q)|=|p-q|$ for any $p,q\in\R^n$.
For any basis $v_1,\dots,v_n$ of $\R^n$, the volumes of $U(v_1,\dots,v_n)$ and $U(f(v_1),\dots,f(v_n))$ have the same absolute non-zero value.
If these volumes are equal, the isometry $f$ is \emph{orientation-preserving}, otherwise $f$ is \emph{orientation-reversing}.
Any orientation-preserving isometry $f$ is a composition of translations and rotations, and can be included into a continuous family of isometries $f_t$, where $t\in[0,1]$, $f_0$ is the identity map and $f_1=f$, which is also called a \emph{rigid motion}.
Any orientation-reversing isometry is a composition of a rigid motion and a single reflection in a linear subspace of dimension $n-1$. 
\bs
\end{dfn}

The Voronoi domain defined below is also called the \emph{Wigner-Seitz cell}, \emph{Brillouin zone} or \emph{Dirichlet cell}.
We use the word \emph{domain} to avoid a confusion with a unit cell, which is a parallelepiped spanned by a vector basis.
Though the Voronoi domain can be defined for any point of a lattice, it will suffice to consider only the origin $0$.

\begin{dfn}[Voronoi domain and Voronoi vectors of a lattice]
\label{dfn:Voronoi_vectors}
The \emph{Voronoi domain} of a lattice $\La$ is the neighbourhood $V(\La)=\{p\in\R^n: |p|\leq|p-v| \text{ for any }v\in\La\}$ of the origin $0\in\La$ consisting of all points $p$ that are non-strictly closer to $0$ than to other points $v\in\La$.
A vector $v\in\La$ is called a \emph{Voronoi vector} if the bisector hyperspace $H(0,v)=\{p\in\R^n \vl p\cdot v=\frac{1}{2}v^2\}$ between 0 and $v$ intersects $V(\La)$.
If $V(\La)\cap H(0,v)$ is an $(n-1)$-dimensional face of $V(\La)$, then $v$ is called a \emph{strict} Voronoi vector. 
\bs
\end{dfn}


Theorem~\ref{thm:reduction} proves that any lattice in $\R^3$ has an obtuse superbase of vectors whose pairwise scalar products are non-positive and are called \emph{Selling parameters}.
For any superbase in $\R^n$, the opposite parameters $p_{ij}=-v_i\cdot v_j$ can be interpreted as conorms of lattice characters, functions $\chi: \La\to\{\pm 1\}$ satisfying $\chi(u+v)=\chi(u)\chi(v)$), see \citeasnoun[Theorem~6]{conway1992low}.
Hence $p_{ij}$ will be shortly called \emph{conorms}. 

\begin{dfn}[obtuse superbase and its conorms $p_{ij}$]
\label{dfn:conorms}
For any basis $v_1,\dots,v_n$ in $\R^n$, the \emph{superbase} $v_0,v_1,\dots,v_n$ includes the vector $v_0=-\sum\limits_{i=1}^n v_i$.
The \emph{conorms} $p_{ij}=-v_i\cdot v_j$ are equal to the negative scalar products of the vectors above. 
The superbase is called \emph{obtuse} if all conorms $p_{ij}\geq 0$, so all angles between vectors $v_i,v_j$ are non-acute for distinct indices $i,j\in\{0,1,\dots,n\}$.
The superbase is called \emph{strict} if all $p_{ij}>0$.
\bs
\end{dfn}

\citeasnoun[formula (1)]{conway1992low} has a typo initially defining $p_{ij}$ as exact Selling parameters, but their Theorems 3,7,8 explicitly use non-negative $p_{ij}=-v_i\cdot v_j\geq 0$.
\medskip

The indices of a conorm $p_{ij}$ are distinct and unordered, so we assume that $p_{ij}=p_{ji}$.
A 1D lattice generated by a vector $v_1$ has the obtuse superbase of $v_0=-v_1$ and $v_1$, so the only conorm $p_{01}=-v_0\cdot v_1=v_1^2$ is the squared norm of $v_1$.
Any basis of $\R^n$ has $\dfrac{n(n+1)}{2}$ conorms $p_{ij}$, for example three conorms $p_{01},p_{02},p_{12}$ in dimension 2.
 
Definition~\ref{dfn:partial_sums} introduces partial sums $v_S$ for any superbase $\{v_i\}_{i=0}^n$ of a lattice $\La$.

\begin{dfn}[partial sums $v_S$ and their vonorms]
\label{dfn:partial_sums}
Let a lattice $\La\subset\R^n$ have any superbase $v_0,v_1,\dots,v_n$ with $v_0=-\sum\limits_{i=1}^n v_i$. 
For any proper subset $S\subset\{0,1,\dots,n\}$ of indices,
 consider its complement $\bar S=\{0,1,\dots,n\}-S$ and the \emph{partial sum} $v_S=\sum\limits_{i\in S} v_i$ whose squared lengths $v_S^2$ are called \emph{vonorms} of the superbase $\{v_i\}_{i=0}^n$.
The \emph{vonorms} can be expressed as $v_S^2=(\sum\limits_{i\in S} v_i)(-\sum\limits_{j\in\bar S}v_j)=-\sum\limits_{i\in S,j\in\bar S}v_{j}\cdot v_j=\sum\limits_{i\in S,j\in\bar S}p_{ij}$.
\bs
\end{dfn}

\cite{conway1992low} call lattices $\La\subset\R^n$ that have an obtuse superbase \emph{lattices of Voronoi's first kind}, which are all lattices in dimensions 2 and 3 by Theorem~\ref{thm:reduction}.

\begin{thm}[obtuse superbase existence]
\label{thm:reduction}
Any lattice $\La$ in dimensions $n=2,3$ has an obtuse superbase  $v_0,v_1,\dots,v_n$ so that $p_{ij}=-v_i\cdot v_j\geq 0$ for any $i\neq j$.
\bs
\end{thm}

Section~7 in \cite{conway1992low} tried to prove Theorem~\ref{thm:reduction} for $n=3$ by example, which turned out to be wrong, see corrections in Fig.~\ref{fig:forms3d_reduction}.
This above will be proved in section~\ref{sec:invariants3d} by reducing a basis to an obtuse superbase and correcting key details from pages 60-63 in \cite{conway1992low}.
\medskip

Lemma~\ref{lem:partial_sums} will later help to prove that a lattice is uniquely determined up to isometry by an obtuse superbase, hence by its vonorms or, equivalently, conorms.

\begin{lem}[Voronoi vectors $v_S$, Theorem~3 in \cite{conway1992low}]
\label{lem:partial_sums}
For any obtuse superbase $v_0,v_1,\dots,v_n$ of a lattice, all partial sums $v_S$ from Definition~\ref{dfn:partial_sums} split into $2^n-1$ symmetric pairs $v_S=-v_{\bar S}$, which are Voronoi vectors representing distinct $2\La$-classes in $\La/2\La$.
All Voronoi vectors $v_S$ are strict if and only if all $p_{ij}>0$.
\bs
\end{lem}

\section{Voforms and coforms of an obtuse superbase of a lattice in dimension 3}
\label{sec:forms3d}

\begin{figure}
\label{fig:forms3d}
\caption{\textbf{Left}: the Fano plane is a set of seven nodes arranged in triples shown by six lines and one circle.
\textbf{Middle}: nodes of the voform $\VF(\La)$ are marked by vonorms $v_i^2$ and $v_{ij}^2$.
\textbf{Right}: nodes of the coform $\CF(\La)$ are marked by conorms $p_{ij}$ and 0.}
\includegraphics[height=31mm]{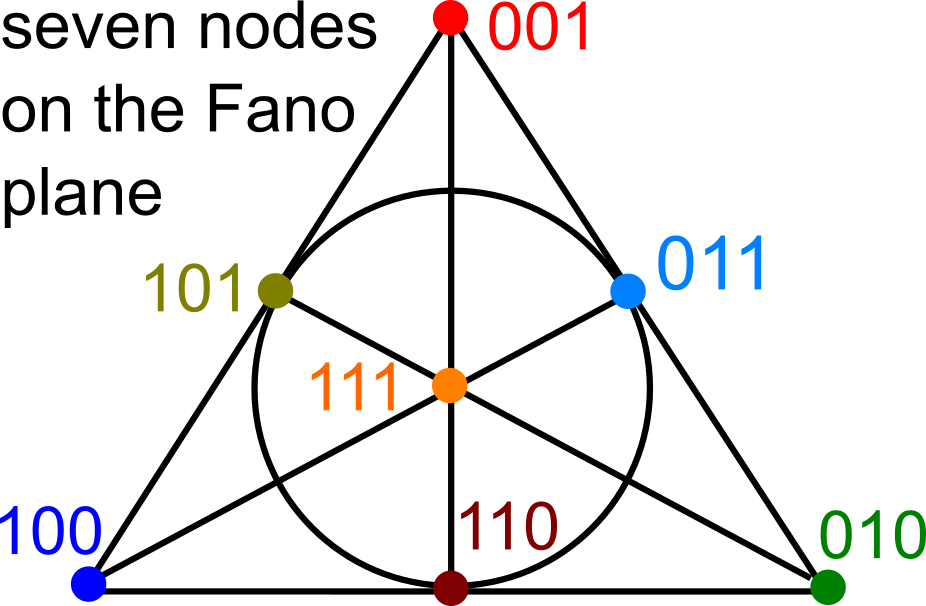}
\hspace*{0mm}
\includegraphics[height=31mm]{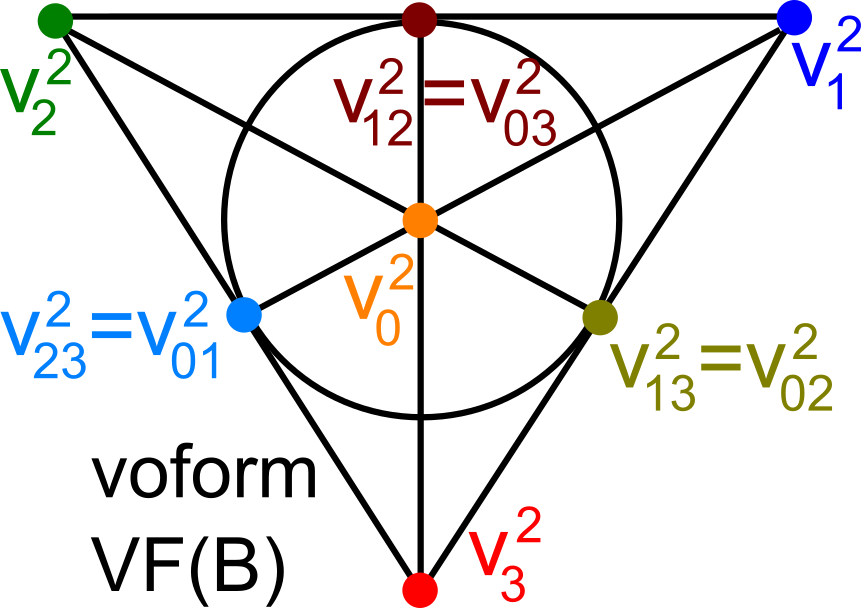}
\hspace*{0mm}
\includegraphics[height=31mm]{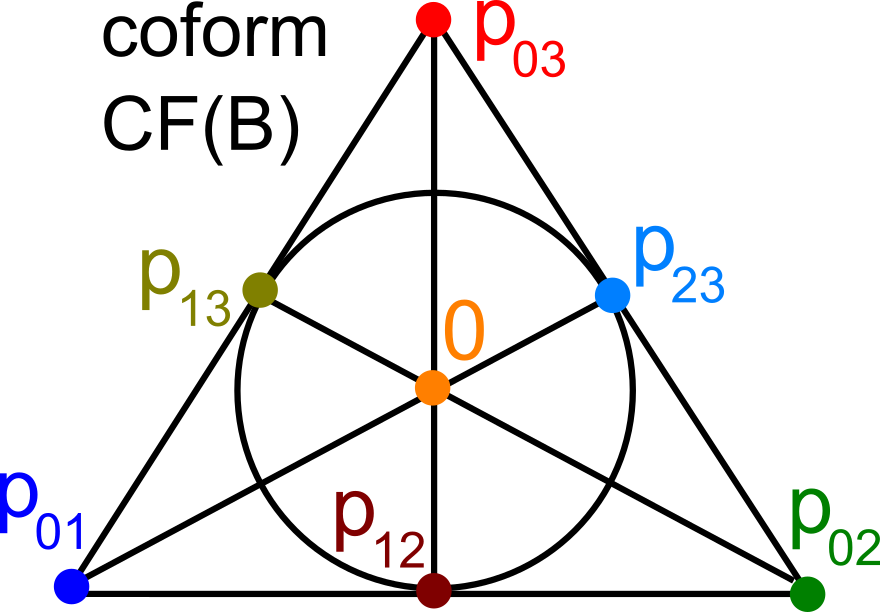}
\end{figure}

For a lattice $\La\subset\R^3$ with an obtuse superbase $B$, Definition~\ref{dfn:forms3d} introduces the voform $\VF(B)$ and the coform $\CF(B)$.
These forms are Fano planes marked by vonorms and conorms, respectively.
The \emph{Fano} projective plane of order 2 consists of seven non-zero classes (called \emph{nodes}) of the space $\La/2\La$, arranged in seven triples (called \emph{lines}).
If we mark these nodes by 3-digit binary numbers $001$, $010$, $011$, $100$, $101$, $110$, $111$, the digit-wise sum of any two numbers in each line equals the third number modulo 2, see Fig.~\ref{fig:forms3d}.
Lemma~\ref{lem:forms3d_invariants} will justify that $\VF,\CF$ are well-defined for any lattice $\La\subset\R^3$.

\begin{dfn}[voform and coform of an obtuse superbase]
\label{dfn:forms3d} 
Any obtuse superbase $B=(v_0,v_1,v_2,v_3)$ in $\R^3$ has seven pairs of partial sums $\pm v_0=\mp(v_1+v_2+v_3)$, $\pm v_1$, $\pm v_2$, $\pm v_3$, $\pm(v_1+v_2)$, $\pm(v_1+v_3)$, $\pm(v_2+v_3)$. 
Definition~\ref{dfn:partial_sums} expresses their vonorms as
$v_i^2=p_{ij}+p_{ik}+p_{il}$ for the unordered triple $\{j,k,l\}=\{0,1,2,3\}-\{i\}$, for instance $v_0^2=p_{01}+p_{02}+p_{03}$.
Similarly, $v_{ij}^2=(v_i+v_j)^2=(-v_k-v_l)^2=p_{ik}+p_{il}+p_{jk}+p_{jl}$ for the unordered pair $\{k,l\}=\{0,1,2,3\}-\{i,j\}$.
The seven vonorms above have the linear relation
$v_0^2+v_1^2+v_2^2+v_3^2=v_{01}^2+v_{02}^2+v_{03}^2$.
The six conorms are conversely expressed as $p_{ij}=\dfrac{1}{2}(v_i^2+v_j^2-v_{ij}^2)$ for any distinct indices $i,j\in\{0,1,2,3\}$.
\medskip

The \emph{voform} $\VF(B)$ is the Fano plane in Fig.~\ref{fig:forms3d} with four nodes marked by $v_0^2,v_1^2,v_2^2,v_3^2$ and three nodes marked by $v_{12}^2, v_{23}^2,v_{13}^2$ so that $v_0^2$ is in the centre, $v_1^2$ is opposite to $v_{23}^2$, etc.
The \emph{coform} $\CF(B)$ is the dual Fano plane in Fig.~\ref{fig:forms3d} with three nodes marked by $p_{12},p_{23},p_{13}$ and three nodes marked by $p_{01},p_{02},p_{03}$, the centre is marked by $0$.
\bs
\end{dfn}

The \emph{zero} conorm $p_0=0$ at the centre of the coform $\CF(B)$ seems mysterious, because \cite{conway1992low} gave no formula for $p_0$, which also wrongly became non-zero in their Fig.~5.
This past mystery is explained by Lemma~\ref{lem:vonorms<->conorms}.
The proof of Theorem~\ref{thm:reduction} for $n=3$ will correct more details in \citeasnoun[Fig.~5]{conway1992low}.

\begin{lem}[6 conorms $\lra$ 7 vonorms]
\label{lem:vonorms<->conorms}
For any distinct indices $i,j\in\{0,1,2,3\}$, the conorm $p_{ij}$ in $\CF(B)$ of any superbase $B$ defines the dual line in the voform $\VF(B)$ through the nodes marked by $v_{ij}^2,v_k^2,v_l^2$ for $\{k,l\}=\{0,1,2,3\}-\{i,j\}$. 
Then
$$4p_{ij}=v_i^2+v_j^2+v_{ik}^2+v_{jk}^2-v_{ij}^2-v_k^2-v_l^2, \leqno{(\ref{lem:vonorms<->conorms}a)}$$
where the vonorms with negative signs are in the line of the voform $\VF(B)$ dual to $p_{ij}$.
The zero conorm $p_0=0$ in $\CF(B)$ can be computed by the similar formula
$$4p_{0}=v_0^2+v_1^2+v_2^2+v_3^2-v_{01}^2-v_{02}^2-v_{03}^2=0, \leqno{(\ref{lem:vonorms<->conorms}b)}$$
where the line dual to the zero conorm $p_0$ is the `circle' through $v_{01}^2,v_{02}^2,v_{03}^2$.
\bs
\end{lem}
\begin{proof}
Since all indices $i,j,k,l\in\{0,1,2,3\}$ are distinct, formula (\ref{lem:vonorms<->conorms}a) is symmetric in $k,l$ due to $v_{ik}^2+v_{jk}^2=v_{il}^2+v_{jl}^2$ following from $v_{ik}=v_i+v_k=-(v_j+v_l)=-v_{jl}$ and $v_{jk}=v_j+v_k=-(v_i+v_l)=-v_{il}$.
To prove (\ref{lem:vonorms<->conorms}a), we simplify its right hand side:
$$v_i^2+v_j^2+v_{ik}^2+v_{jk}^2-v_{ij}^2-v_k^2-v_l^2
=v_i^2+v_j^2+(v_i+v_k)^2+(v_j+v_k)^2-(v_i+v_j)^2-v_k^2-$$
$$-(-v_i-v_j-v_k)^2
=v_i^2+v_j^2+(v_i^2+2v_iv_k+v_k^2)+(v_j^2+2v_jv_k+v_k^2)-$$
$$-(v_i^2+2v_iv_j+v_j^2)-v_k^2-(v_i^2+v_j^2+v_k^2+2v_i v_j+2v_iv_k+2v_jv_k)=-4v_iv_j=4p_{ij}
.$$
Formula~(\ref{lem:vonorms<->conorms}b) follows from
$v_0^2+v_1^2+v_2^2+v_3^2=v_{01}^2+v_{02}^2+v_{03}^2$
in Definition~\ref{dfn:forms3d}.
\end{proof}
 
\begin{figure}
\label{fig:forms3d_permutations}
\caption{Actions of permutations $1\lra 2$ and $0\lra 1$ on voforms (top) and coforms.}
\includegraphics[width=\textwidth]{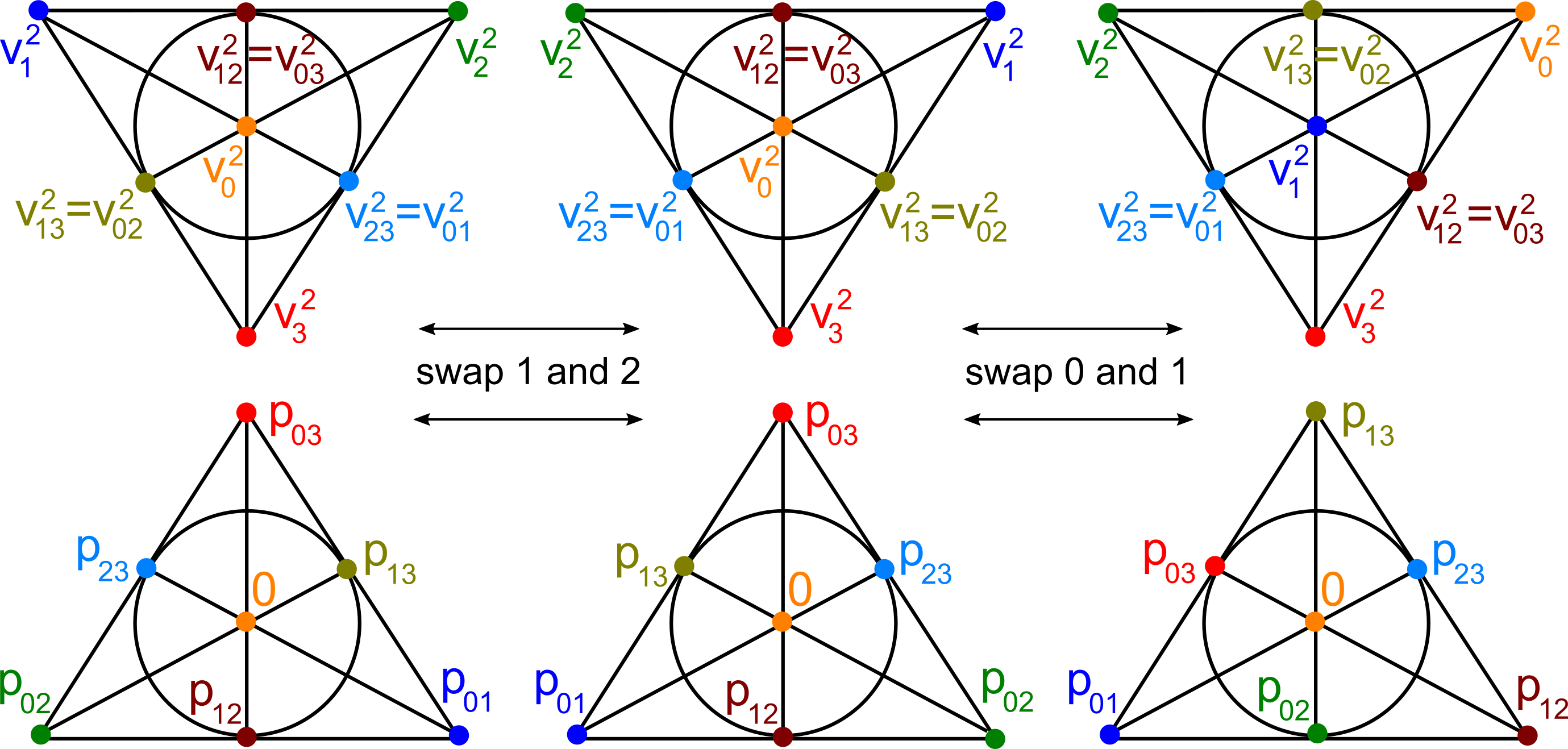}
\end{figure}

\begin{dfn}[isomorphisms of voforms and coforms]
\label{dfn:isomorphisms3d}
An \emph{isomorphism} of voforms is a permutation $\si\in S_4$ of indices $0,1,2,3$, which maps vonorms as follows: $v_i^2\mapsto v_{\si(i)}^2$, $v_{ij}^2\mapsto v_{\si(i)\si(j)}^2$, where $v_{ij}^2=v_{ji}^2$.
If we swap $v_1^2,v_2^2$, then we also swap only $v_{13}^2=v_{02}^2$ and $v_{23}^2=v_{01}^2$.
If we swap $v_0^2,v_1^2$, then we also swap only $v_{12}^2=v_{03}^2$ and $v_{02}^2=v_{13}^2$, see Fig.~\ref{fig:forms3d_permutations}.
An \emph{isomorphism} of coforms is a permutation $\si\in S_4$ of $0,1,2,3$, which maps conorms as follows: $p_{ij}\mapsto p_{\si(i)\si(j)}$, where $p_{ij}=p_{ji}$.
An isomorphism above is called \emph{orientation-preserving} if the permutation $\si$ of the indices $0,1,2,3$ is even (or positive) meaning that $\si$ decomposes into an even number of transpositions $i\lra j$.
\bs
\end{dfn}

\begin{exa}[voforms and coforms as matrices]
\label{exa:forms3dmat}
Any voform  can be written as the $2\times 3$ matrix $\VF(\La)=\mat{v_{23}^2}{v_{13}^2}{v_{12}^2}{v_1^2}{v_2^2}{v_3^2}$, where $v_{23}^2=v_{01}^2$ is above $v_{1}^2$ and so on.
The 7th vonorm can be found as $v_0^2= v_{23}^2 + v_{13}^2 + v_{12}^2  - v_1^2 - v_2^2- v_3^2$ and is unnecessary to include.
Similarly, any coform can be written as 
$\CF(\La)=\mat{p_{23}}{p_{13}}{p_{12}}{p_{01}}{p_{02}}{p_{03}}$.
\medskip

The permutations $1\lra 2$ and $0\lra 1$ affect the voforms and coforms as follows:
$$\mat{
\color[rgb]{0,0.5,0.5}{v_{13}^2}}{
\color[rgb]{0,0.5,1}{v_{23}^2}}{
v_{12}^2}{
\color[rgb]{0,0.5,0}{v_2^2}}{
\color[rgb]{0,0,1}{v_1^2}}{
v_3^2}
\stackrel{1\lra 2}{\longleftrightarrow}\VF(\La)=
\mat{
\color[rgb]{0,0.5,1}{v_{23}^2}}{
\color[rgb]{0,0.5,0.5}{v_{13}^2}}{
\color[rgb]{0.5,0,0}{v_{12}^2}}{
\color[rgb]{0,0,1}{v_1^2}}{
\color[rgb]{0,0.5,0}{v_2^2}}{
v_3^2}
\stackrel{0\lra 1}{\longleftrightarrow}
\mat{
v_{23}^2}{
\color[rgb]{0.5,0,0}{v_{12}^2}}{
\color[rgb]{0,0.5,0.5}{v_{13}^2}}{
\color[rgb]{1,0.5,0}{v_0^2}}{
v_2^2}{
v_3^2}
\leqno{(\ref{exa:forms3dmat}a)}$$
$$\mat{
\color[rgb]{0.5,0.5,0}{p_{13}}}{
\color[rgb]{0,0.5,1}{p_{23}}}{
p_{12}}{
\color[rgb]{0,0.5,0}{p_{02}}}{
\color[rgb]{0,0,1}{p_{01}}}{
p_{03}}
\stackrel{1\lra 2}{\longleftrightarrow}\CF(\La)=
\mat{
\color[rgb]{0,0.5,1}{p_{23}}}{
\color[rgb]{0.5,0.5,0}{p_{13}}}{
\color[rgb]{0.5,0,0}{p_{12}}}{
\color[rgb]{0,0,1}{p_{01}}}{
\color[rgb]{0,0.5,0}{p_{02}}}{
\color[rgb]{1,0,0}{p_{03}}}
\stackrel{0\lra 1}{\longleftrightarrow}
\mat{
p_{23}}{
\color[rgb]{1,0,0}{p_{03}}}{
\color[rgb]{0,0.5,0}{p_{02}}}{
p_{01}}{
\color[rgb]{0.5,0,0}{p_{12}}}{
\color[rgb]{0.5,0.5,0}{p_{13}}}
\leqno{(\ref{exa:forms3dmat}b)}$$

Since an action on the voform may require the 7th vonorm $v_0^2$, we will mainly use conorms for classifying lattices and defining metrics on their isomery classes.
\medskip

In general, any transposition of non-zero indices $i\lra j$ swaps the columns $i$ and $j$ in $\CF(\La)$.
Any transposition $0\lra i$ for $i\neq 0$ diagonally swaps two pairs in the columns different from $i$.
In all cases, two conorms from one column remain in one column.
\bs
\end{exa}

Permutations~(\ref{exa:forms3dmat}ab) show that coforms of six conorms are easier than voforms, which essential require seven vonorms since $v_0^2$ appears after the transposition $0\lra 1$. 
  
\begin{figure}
\label{fig:DC7examples}
\caption{The lattices defined by non-isomorphic coforms $\CF(\La)\not\sim\CF(\La')$ are not isometric but the origin $0$ has the same distances to its seven closest neighbours.}
\includegraphics[width=\textwidth]{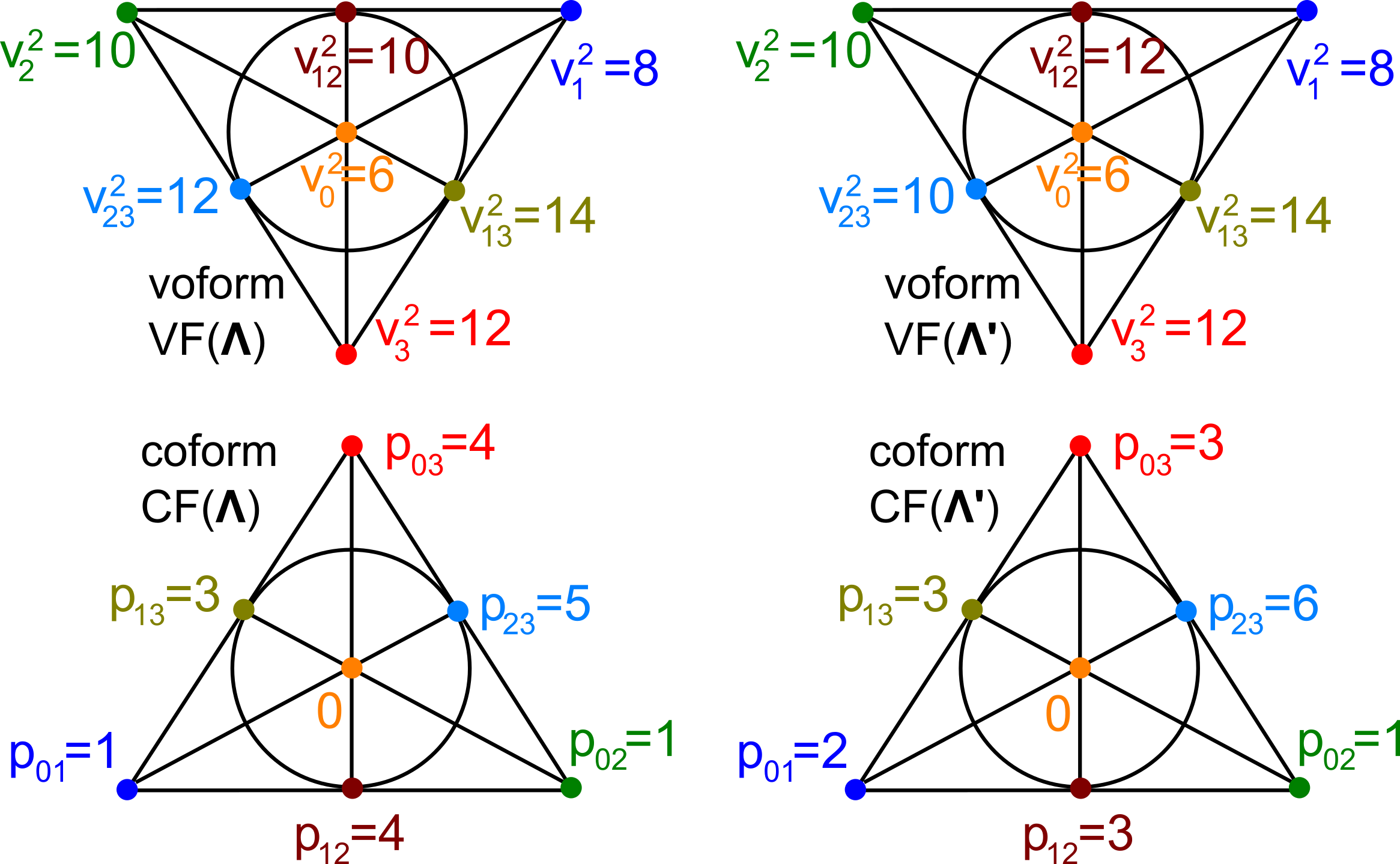}
\end{figure}

\begin{exa}[non-isometric lattices with $DC^7(\La,\La')=0$]
\label{exa:dc7=0}
Fig.~\ref{fig:DC7examples} shows that we can not arbitrarily permute conorms or vonorms without changing our lattice.
Only $4!=24$ permutations are allowed for isomorphisms in Definition~\ref{dfn:forms3d}.
The voforms in Fig.~\ref{fig:DC7examples} differ by a single transposition $10\lra 12$ for the vonorms $v_{12}^2$ and $v_{23}^2$.
The coforms in Fig.~\ref{fig:DC7examples} are computed from the voforms by the formulae in Definition~\ref{dfn:forms3d}.
Since coforms consist of different numbers, they are not isomorphic and will give rise to non-isometric lattices $\La,\La'$, see an explicit reconstruction in Lemma~\ref{lem:superbase_reconstruction}. 
\medskip

In these lattices $\La,\La'\subset\R^3$ the origin $0$ has the same distances $|v_0|$, $|v_1|$, $|v_2|$, $|v_3|$, $|v_{12}|$, $|v_{23}|$, $|v_{13}|$ to its seven closest Voronoi neighbours.  
Hence the $DC^7$ functions taking the Euclidean distance between these 7-dimensional distance vectors \cite{andrews2019space} vanishes
for $\La,\La'$.
Our colleagues Larry Andrews and Herbert Bernstein quickly checked that $\La,\La'$ can be distinguished by the 8th distance from the origin to its 8th closest neighbour.   
However, the example Fig.~\ref{fig:DC7examples} can be extended to tan infinite 6-parameter family of pairs $\La,\La'$ with $DC^7(\La,\La')=0$ as follows.
\medskip

Add an arbitrary coform of any conorms $q_{ij}\geq 0$ to $\CF(\La),\CF(\La)$ `conorm-wise'.
Definition~\ref{dfn:forms3d} implies that the voforms consist of the same 7 numbers, e.g.
$$\La:\quad v_0^2=(p_{01}+q_{01})+(p_{02}+q_{02})+(p_{03}+q_{03})
=1+4+1+q_{01}+q_{02}+q_{03},$$
$$\La':\quad v_0^2=(p'_{01}+q_{01})+(p'_{02}+q_{02})+(p'_{03}+q_{03})
=2+1+3+q_{01}+q_{02}+q_{03}.$$
The coforms will remain non-isomorphic if we exclude the singular case
when $q_{23}+q_{01} = q_{12}+q_{03}$.
These 6-parameter family of non-isometric lattices $\La,\La'$ might be distinguished by 8 or more distances from the origin to its neighbours, but this conclusion requires a theoretical argument. 
The root metric in Definition~\ref{dfn:metrics3d} will provably satisfy the first metric axiom: $\RM_d(\La,\La')=0$ if and only if $\La,\La'$ are isometric.
\bs
\end{exa}
 
\section{Unique root forms are isometry invariants of lattices in dimension 3}
\label{sec:invariants3d}

Isomorphisms from Definition~\ref{dfn:isomorphisms3d} help unambiguously order the six conorms within a coform and define a unique root form, which will classify lattices up to isometry.

\begin{dfn}[the root form $\RF(\La)$ of an obtuse superbase]
\label{dfn:root_form3d}
Since any obtuse superbase $B$ has only non-negative conorms, the six \emph{root products} $r_{ij}=\sqrt{p_{ij}}$ are well-defined for all distinct indices $i,j\in\{0,1,2,3\}$ and have the same units as original coordinates of basis vectors, for example in Angstroms: $1\AA=10^{-10}$m.
\medskip

For any matrix of root products
$\mat{r_{23}}{r_{13}}{r_{12}}{r_{01}}{r_{02}}{r_{03}}$,
a permutation of indices 1,2,3 as in (\ref{exa:forms3dmat}a) allows us to arrange the three columns in any order.
The composition of transpositions $0\lra i$ and $j\lra k$ for distinct $i,j,k\neq 0$ vertically swaps the root products in columns $j$ and $k$, for example apply the transposition $2\lra 3$ to the result of $0\lra 1$ in (\ref{exa:forms3dmat}b).
So we can put the minimum value $r_{min}$ into the top left position ($r_{23}$).
Then we consider the four root products in columns 2 and 3.
Keeping column 1 fixed, we can put the minimum of these four into the top middle position ($r_{13}$).
Then the resulting root products in the top row should be in increasing order.
\medskip

If the top left and top middle root products are accidentally equal ($r_{23}=r_{13}$), we can put their counterparts ($r_{01}$ and $r_{02}$) in the bottom row of columns 1,2 in increasing order.  
If the top middle and top right root products are accidentally equal ($r_{13}=r_{12}$), we can put their counterparts ($r_{02}$ and $r_{03}$) in the bottom row of columns 2 and 3 in increasing order.  
The resulting matrix is called the \emph{root form} $\RF(B)$ and can be  visualised as in the last picture of Fig.~\ref{fig:forms3d} with root products instead of conorms.
\medskip

For orientation-preserving isomorphism, we have only 12 available permutations of 0,1,2,3 from the group $A_4$ such as the cyclic permutations of the three columns and vertical swaps in two columns, for example realised by the composition of $0\lra 1$ and $2\lra 3$.
These positive permutations still allow us to put the minimum of the six root products into the top left position.
The top row can not be put in increasing order if $r_{13}>r_{12}$ and $r_{02}>r_{03}$.
The vertical swap in columns 2 and 3 can put $(r_{13},r_{12})$, $(r_{02},r_{03})$ in the lexicographic order so that $r_{13}<r_{02}$, if $r_{13}=r_{02}$ then $r_{12}\leq r_{03}$.
\medskip

The only unresolved ambiguity may appear in the case when all root products in the top row equal the minimum value $r_{min}$ of all six.
Then we put the minimum of three remaining root products at the left position in the bottom row.
If five root products equal the minimum value $r_{min}$, the 6th one can be put in the bottom right position.
We got a unique \emph{root form} $\RF^+(\La)$ up to orientation-preserving isomorphism. 
\bs
\end{dfn}

Geometrically, any root product $r_{ij}$ measures non-orthogonality of vectors $v_i,v_j$.
 
\begin{lem}[equivalence of $\VF,\CF,\RF$]
\label{lem:forms3d_equiv}
For any obtuse superbase $B$ in $\R^3$, its voform $\VF(B)$, coform $\CF(B)$ and unique $\RF(B)$ are reconstructible from each other.
\end{lem}
\begin{proof}
The six conorms $p_{ij}$ are uniquely expressed via the seven vonorms $v_i^2,v_{ij}^2$ by formulae (\ref{dfn:forms3d}ab) and vice versa.
If we apply a permutation of indices $0,1,2,3$ to the conorms, the same permutation applies to the vonorms.
Hence we have a 1-1 bijection $\CF(\La)\lra\VF(\La)$ up to (orientation-preserving) isomorphism. 
The root form $\RF(\La)$ is uniquely defined by ordering root products without any need for isomorphisms.
\end{proof}

\begin{lem}[isometry$\to$isomorphism]
\label{lem:forms3d_invariants}
Any (orientation-preserving) isometry of obtuse superbases $B\to B'$ induces an (orientation-preserving, respectively) isomorphism of voforms $\VF(B)\sim\VF(B')$, coforms $\CF(B)\sim\CF(B')$ and keeps $\RF(B)=\RF(B')$.
\bs
\end{lem}
\begin{proof}
Any isometry preserves lengths and scalar products of vectors.
\end{proof}

Lemma~\ref{lem:reduction} will help find an obtuse superbase for any lattice $\La\subset\R^3$.

\begin{lem}[reduction]
\label{lem:reduction}
Let $B=(v_0,v_1,v_2,v_3)$ be any superbase of a lattice $\La\subset\R^3$.
For any distinct $i,j,k,l\in\{0,1,2,3\}$, let the new superbase vectors be 
$u_i=-v_i$, $u_j=v_j$, $u_k=v_{ik}=v_i+v_k$, $u_l=v_{il}=v_i+v_j$.
Then all vonorms remain the same or swap their places, and the only change is $u_{ij}^2=v_{ij}^2-4\ep$, where $\ep=v_i\cdot v_j$.
The conorms $q_{\bullet}$ of the new vectors $u_{\bullet}$ are updated 
as in Fig.~\ref{fig:forms3d_reduction} for $(i,j)=(1,3)$, $(k,l)=(0,2)$.
$$q_{ij}=\ep,\; 
q_{jk}=p_{jk}-\ep,\; 
q_{jl}=p_{jl}-\ep,\; 
q_{ik}=p_{il}-\ep,\; 
q_{il}=p_{ik}-\ep,\; 
q_{kl}=p_{kl}+\ep. \quad \blacksquare
\leqno{(\ref{lem:reduction})}$$
\end{lem}
\begin{proof}
If initial vectors $v_{\bullet}$ form a superbase, which means that $v_i+v_j+v_k+v_l=0$, then so do the new vectors: $u_i+u_j+u_k+u_l=(-v_i)+v_j+(v_i+v_k)+(v_i+v_l)=0$.

\begin{figure}
\label{fig:forms3d_reduction}
\caption{Lemma~\ref{lem:reduction} for $i=1$, $k=2$, $j=3$, $l=0$ says that
the new superbase $u_1=-v_1$, $u_2=v_{12}$, $u_3=v_3$, $u_0=v_{01}$ has the new voform $\VF$ and coform $\CF$ shown above.}
\includegraphics[width=\textwidth]{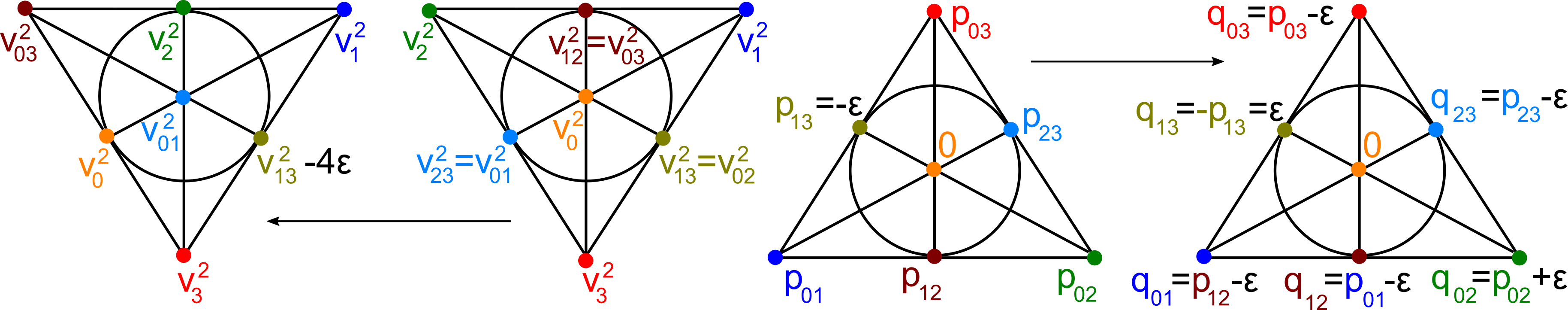}
\end{figure}

For the new superbase $u_i=-v_i$, $u_j=v_j$, $u_k=v_{ik}$, $u_l=v_{il}$,
two vonorms remain the same: $u_i^2=v_i^2$ and $u_j^2=v_j^2$.
Two pairs of vonorms swap their places: $u_k^2=v_{ik}^2$, $u_{ik}^2=(u_i+u_k)^2=v_k^2$ and $u_l^2=v_{il}^2$, $u_{il}^2=(u_i+u_l)^2=v_l^2$.
The final vonorm is
$$u_{ij}^2=u_{kl}^2=(v_j-v_i)^2=(v_i+v_j)^2-4v_i\cdot v_j=v_{ij}^2+4p_{ij}=v_{ij}^2-4\ep, \text{ see Fig.~\ref{fig:forms3d_reduction}.}$$
We similarly check formulae (\ref{lem:reduction}) illustrated in Fig.~\ref{fig:forms3d_reduction} for $i=1$, $k=2$, $j=3$, $l=0$.

\noindent
$q_{ij}=-u_i\cdot u_j=v_i\cdot v_j=-p_{ij}=\ep$

\noindent
$q_{jk}=-u_j\cdot u_k=-v_j\cdot(v_i+v_k)=-v_i\cdot v_j-v_j\cdot v_k=p_{jk}-\ep$

\noindent
$q_{jl}=-u_j\cdot u_l=-v_j\cdot(v_i+v_l)=-v_i\cdot v_j-v_j\cdot v_l=p_{jl}-\ep$

\noindent
$q_{ik}=-u_i\cdot u_k=v_i\cdot(v_i+v_k)=v_i\cdot(-v_j-v_l)=-v_i\cdot v_l-v_i\cdot v_j=p_{il}-\ep$

\noindent
$q_{il}=-u_i\cdot u_l=v_i\cdot(v_i+v_l)=v_i\cdot(-v_j-v_k)=-v_i\cdot v_k-v_i\cdot v_j=p_{ik}-\ep$

\noindent
$q_{kl}=-u_k\cdot u_l=(v_i+v_k)\cdot(v_i+v_l)=v_i\cdot(-v_i-v_j-v_k)-v_k\cdot v_l
=v_i\cdot v_j+p_{kl}
=p_{kl}+\ep$.
\medskip

Notice that the conorm $p_0$ at the centre of $\CF$ remains zero by formula~(\ref{lem:vonorms<->conorms}b):
$$4p_{0}=u_i^2+u_j^2+u_k^2+u_l^2-u_{ij}^2-u_{ik}^2-u_{il}^2
=v_i^2+v_j^2+(v_i+v_k)^2+(v_i+v_l)^2-(v_j-v_i)^2-v_k^2-v_l^2=$$
$$=v_i^2+v_j^2+(v_i^2+2v_iv_k+v_k^2)+(v_j+v_k)^2-(v_i^2-2v_iv_j+v_j^2)-v_k^2-(v_i+v_j+v_k)^2=$$
$$=v_i^2+v_j^2+v_k^2+2v_iv_j+2v_jv_k+2v_iv_k-(v_i+v_j+v_k)^2
=0.$$
Hence all central conorms $p_0$ in \citeasnoun[Fig.~5]{conway1992low} should be 0.
\end{proof}

\begin{proof}[Proof of Theorem~\ref{thm:reduction} for $n=3$]
We will reduce any superbase $B=(v_0,v_1,v_2,v_3)$ of a lattice $\La\subset\R^3$ to make all conorms $p_{ij}$ non-negative.
Starting from any negative conorm $p_{ij}=-\ep<0$, we change the superbase by Lemma~\ref{lem:reduction}.
This reduction leads to the positive conorm $q_{ij}=\ep$, not zero as in \citeasnoun[Fig.~4(b)]{conway1992low}.
\medskip

Four other conorms decrease by $\ep>0$ and can potentially become negative, which requires a new reduction by Lemma~\ref{lem:reduction} and so on.
To prove that the reduction process always finishes, notice that six vonorms keep or swap their values, but one vonorm always decreases by $4\ep>0$.
Every reduction can make superbase vectors only shorter, but not shorter than a minimum distance between points of $\La$.
The angle between $v_i,v_j$ can have only finitely many values when lengths of $v_i,v_j$ are bounded.
Hence the scalar product $\ep=v_i\cdot v_j>0$ cannot converge to 0.
Since every reduction makes one partial sum $v_S$ shorter by a positive constant, while other six vectors $v_S$ keep or swap their lengths, the reductions by Lemma~\ref{lem:reduction} should finish in finitely many steps.
\end{proof}

A reduction of lattice bases for real crystals has many efficient implementations.
Theoretical estimates for reduction steps are discussed in \cite{nguyen2009low}.
\medskip

\begin{lem}
\label{lem:isometric_superbases} 
All obtuse superbases of any lattice $\La\subset\R^3$ are isometric.
Hence $\VF(\La)$, $\CF(\La)$, $\RF(\La)$ are independent of a superbase (well-defined up to isomorphism).
\bs
\end{lem}
\begin{proof}
By Lemma~\ref{lem:partial_sums} for $n=3$, if $\La$ has a strict obtuse superbase $v_0,v_1,v_2,v_3$, all Voronoi vectors of $\La$ are 7 pairs of partial sums $\pm v_S$ for the vectors $v_S$ from the list
$$v_0,\quad
v_1,\quad
v_2,\quad
v_3,\quad
v_2+v_3=-(v_0+v_1),\quad
v_3+v_1=-(v_0+v_2),\quad
v_1+v_2=-(v_0+v_3).$$
 
In this generic case, the Voronoi domain $V(\La)$ is a truncated octahedron. 
First, $V(\La)$ has four pairs of opposite hexagonal faces obtained by cutting corners in four pairs of opposite triangular faces in an octahedron.
The normal vectors of these hexagons are the Voronoi vectors $\pm v_i$, $i=0,1,2,3$.
Second, $V(\La)$ has three pairs of opposite parallelogram faces obtained by cutting three pairs of opposite vertices in an octahedron.
The normal vectors of these faces are the Voronoi vectors $v_i+v_j$ for distinct $i,j\in\{0,1,2,3\}$.
Hence a superbase $\{v_0,v_1,v_2,v_3\}$ of any generic $\La$ is determined up to a sign by the four pairs of opposite hexagonal faces in $V(\La)$.
\medskip

If a superbase of $\La$ is non-strict, one conorm vanishes, say $p_{12}=0$, so the basis vectors $v_1,v_2$ become orthogonal.
If $v_1$ or $v_2$ has strictly obtuse angles with both other vectors $v_3$ and $v_0$, there are still only two symmetric superbases $\pm\{v_0,v_1,v_2,v_3\}$.
If (say) $v_1$ becomes orthogonal to both $v_2,v_3$, we get the new pair of symmetric superbases $\pm\{v_1-v_2-v_3,-v_1,v_2,v_3\}$ related to $\pm\{-v_1-v_2-v_3,v_1,v_2,v_3\}$ by the mirror reflection with respect to the plane orthogonal to $v_1$. 
If two more vectors $v_2,v_3$ become orthogonal, the Voronoi domain $V(L)$ is a rectangular box with four pairs of symmetric superbases, which are all related by mirror reflections in $\R^3$.
\medskip

Any (even) permutation of vectors $v_0,v_1,v_2,v_3$ induces an (orientation-preserving) isomorphism of voforms and coforms and keeps the root form invariant.
Lemma~\ref{lem:forms3d_invariants} implies that $\VF(\La)$, $\CF(\La)$, $\RF(\La)$ are independent of a superbase $B$ of $\La$. 
\end{proof}


\begin{exa}[root forms of orthorhombic lattices]
\label{exa:orthorhombic_lattices}
Scaling any lattice $\La$ by a factor $s\in\R$ multiplies all root products in $\RF(\La)$ by $s$.
The primitive orthorhombic lattice ($oP$) with side lengths $0\leq a\leq b\leq c$ has the obtuse superbase $v_1=(a,0,0)$, $v_2=(0,b,0)$, $v_3=(0,0,c)$, $v_0=(-a,-b,-c)$ and the root form $\RF(oP)=\mat{0}{0}{0}{a}{b}{c}$.
\medskip

Let a Base-centred Orthorhombic lattice ($oS$) have the underlying cube of side lengths $2a\leq 2b\leq 2c$. 
The obtuse superbase $v_1=(2a,0,0)$, $v_2=(-a,b,0)$, $v_3=(0,0,c)$, $v_0=(-a,-b,-c)$ gives the root form $\RF(oS)=\mat{0}{0}{a\sqrt{2}}{a\sqrt{2}}{\sqrt{b^2-a^2}}{c}$, where the first two columns should be swapped if $a\sqrt{2}>\sqrt{b^2-a^2}$ or $3a^2>b^2$.
\medskip

In the above notations, a Face-centred Orthorhombic lattice ($oF$) has the obtuse superbase $v_1=(a,b,0)$, $v_2=(a,-b,0)$, $v_3=(-a,0,c)$, $v_0=(-a,0,-c)$.
If $b^2<2a^2$, the root form is $\RF(oF)=\mat{\sqrt{b^2-a^2}}{a}{a}{\sqrt{c^2-a^2}}{a}{a}$, otherwise the first column should be swapped with the last column.
For a Body-centred Orthorhombic lattice ($oI$) on the same cube above, assume the triangle with side lengths $a,b,c$ is acute to guarantee non-negative conorms.
This lattice has the obtuse superbase $v_1=(a,b,-c)$, $v_2=(a,-b,c)$, $v_3=(-a,b,c)$, $v_0=(-a,-b,-c)$ and the root form $\RF(oI)=\mat{\sqrt{a^2+b^2-c^2}}{\sqrt{a^2-b^2+c^2}}{\sqrt{-a^2+b^2+c^2}}{\sqrt{a^2+b^2-c^2}}{\sqrt{a^2-b^2+c^2}}{\sqrt{-a^2+b^2+c^2}}$, where the root products in each row are in increasing order as expected due to $a\leq b\leq c$.
\bs
\end{exa}

\section{The simpler space of obtuse superbases up to isometry in dimension 3}
\label{sec:superbases/isometry3d}

\begin{dfn}[space $\OSI^{(3)}$ of obtuse superbases up to isometry]
\label{dfn:OSI}
Let $B=\{v_i\}_{i=0}^3$ and $B'=\{u_i\}_{i=0}^3$ be any obtuse superbases in $\R^3$.
The maximum Euclidean length of vector differences $L_{\infty}(B,B')=\min\limits_{R\in\Or(\R^3)}\max\limits_{i=0,1,2,3}|R(u_i)-v_i|$ is minimised over all orthogonal maps $R$ from the compact group $\Or(\R^3)$.
Let $\OSI^{(3)}$ denote the space of all \emph{obtuse superbases up to isometry} in $\R^3$, which we equip with the metric $L_{\infty}$.
For orientation-preserving isometries, we have the space $\OSI^{(3)+}$ with the metric $L_{\infty}^+$ defined by minimising over all 3-dimensional rotations from the group $\SO(\R^3)$.
\bs
\end{dfn}

Theorem~\ref{thm:superbases/isometry} substantially reduces the ambiguity of basis representations due to the 1-1 map $\LISP^{(3)}\to\OSI^{(3)}$.
Any fixed lattice $\La\subset\R^3$ has infinitely many (super)bases but only a few obtuse superbases, maximum eight for rectangular Voronoi domains.

\begin{thm}[lattices up to isometry $\lra$ obtuse superbases up to isometry]
\label{thm:superbases/isometry} 
Lattices in $\R^3$ are isometric if and only if any of their obtuse superbases are isometric.
\bs
\end{thm}
\begin{proof}
Part \emph{only if} ($\Rightarrow$): any isometry $f$ between lattices $\La,\La'$ maps any obtuse superbase $B$ of $\La$ to the obtuse superbase $f(B)$ of $\La'$, which should be isometric to any other obtuse superbase of $\La'$ by Lemma~\ref{lem:isometric_superbases}.  
Part \emph{if} ($\Leftarrow$): any isometry between obtuse superbases of $\La,\La'$ linearly extends to an isometry between the lattices $\La,\La'$.
\end{proof}

\begin{lem}[special lattices and their root forms]
\label{lem:special_forms}
\textbf{(a)} 
If the root form $\RF(\La)$ of a lattice $\La\subset\R^3$ has two equal columns with identical root products, for example $r_{12}=r_{13}$ and $r_{02}=r_{03}$, then $\La$ is a mirror reflection of itself. 
\medskip

\noindent
\textbf{(b)}
If the rows of $\RF(\La)$ coincide, then $\La$ is a Face-centred Orthorhombic lattice.
\bs
\end{lem}
\begin{proof}
\textbf{(a)} 
If $r_{12}=r_{13}$ and $r_{02}=r_{03}$, then the vectors $v_2,v_3$ have the same length by formulae of Definition~\ref{dfn:forms3d}:
$v_2^2=p_{02}+p_{12}+p_{23}=p_{03}+p_{13}+p_{23}=v_3^2.$
Then $v_2,v_3$ are mirror images with respect to their bisector plane $P$.
The identity $p_{02}=p_{03}$ implies that $v_0$ has the same angles with the vectors $v_2,v_3$ of equal lengths, also $v_1$ due to $p_{12}=p_{13}$. 
Then both $v_0,v_1$ belong to the bisector plane $P$ between $v_2,v_3$.
Hence the superbase is invariant under the mirror reflection with respect to $P$.
\medskip

\noindent
\textbf{(b)}
If $p_{01}=p_{23}$, $p_{02}=p_{13}$, $p_{03}=p_{12}$, the formulae of Definition~\ref{dfn:forms3d} imply that the vectors $v_0,v_1,v_2,v_3$ have the same squared length equal to $p_{01}+p_{02}+p_{03}$.
The three other partial sums $v_0+v_i$, $i=1,2,3$, are orthogonal to each other.
Indeed, $$(v_0+v_i)\cdot (v_0+v_j)=v_0^2+v_0\cdot v_i +v_0\cdot v_j+v_i\cdot v_j=(p_{01}+p_{02}+p_{03})-p_{0i}-p_{0j}-p_{ij}=0,$$
because $p_{ij}=p_{0k}$ when all indices $i,j,k\in\{1,2,3\}$ are distinct. 
Hence the vectors $v_0+v_i$ form a non-primitive orthogonal basis of $\La$.
Parameters $a,b,c$ of a Face-centred Orthorhombic lattice ($oF$) can be found from Example~\ref{exa:orthorhombic_lattices}.
\end{proof}

\begin{dfn}[sign of a lattice]
\label{dfn:sign_lattice}
A lattice $\La\subset\R^3$ is called \emph{neutral} (or \emph{achiral})  $\La$ maps to itself under a mirror reflection.
If $\La$ is not neutral, we define its positive/negative sign from the orientation-preserving root form $\RF^+(\La)$ as follows.
\medskip

If the root products in the top row of $\RF^+(\La)$ are in strictly increasing (decreasing) order, then $\La$ is called \emph{positive} (\emph{negative}, respectively).
In the exceptional case when the rows of $\RF^+(\La)$ coincide, $\La$ is neutral by Lemma~\ref{lem:special_forms}(b).
\medskip

If the top row contains two zero root products, say $p_{12}=p_{13}=0$, then the vector $v_1$ is orthogonal to both $v_2,v_3$, hence 
$\La$ can be reflected to itself by $v_1\mapsto -v_1$, so $\La$ is neutral.
If two root products in the top row have the same non-zero value, say $r_{13}=r_{12}>0$, then we compare the root products $r_{02}$ and $r_{03}$ below them: if $r_{02}<r_{03}$ then the lattice $\La$ is called \emph{positive}, if $r_{02}>r_{03}$ then $\La$ is called \emph{negative}.
\medskip

If $r_{12}=r_{13}$ and $r_{02}=r_{03}$, then $\La$ is neutral by Lemma~\ref{lem:special_forms}(a).
\bs
\end{dfn}

\begin{exa}[neutral lattices]
\label{exa:sign_lattice}
All orthorhombic lattices from Example~\ref{exa:orthorhombic_lattices} are neutral, because they have a mirror symmetry, which is also visible in their root forms $\RF^+(\La)$ containing other two zeros in the top row ($oP$ and $oS$) or having identical columns ($oF$) or rows ($oI$).
Any monoclinic lattice $\La$ has a superbase $v_1=(a,0,0)$, $v_2=(b\cos\al,b\sin\al,0)$, $v_3=(0,0,c)$, $v_0=(-a-b\cos\al,-b\sin\al,-c)$, where $a\leq b$ and a non-acute angle $\al$ satisfies $a+b\cos\al\geq 0$.
Then $\La$ is neutral and has the root form $\RF^+(\La)=\mat{0}{0}{\sqrt{-ab\cos\al}}{\sqrt{a^2+ab\cos\al}}{\sqrt{b^2+ab\cos\al}}{c}$. 
\bs
\end{exa}

\section{Unique root forms classify all lattices up to isometry in dimension 3}
\label{sec:classification3d}

\begin{lem}[superbase reconstruction]
\label{lem:superbase_reconstruction}
For any lattice $\La\subset\R^2$, an obtuse superbase $B$ of $\La$ can be reconstructed up to isometry from $\VF(\La)$ or $\CF(\La)$ or $\RF(\La)$.
\bs
\end{lem}
\begin{proof}
Since $\VF(\La),\CF(\La),\RF(\La)$ are expressible via each other by Lemma~\ref{lem:forms3d_equiv}, it suffices to reconstruct an obtuse superbase $v_0,v_1,v_2,v_3$ of $\La$ from $\RF(\La)$.
The positions of root products $r_{ij}=\sqrt{-v_i\cdot v_j}$ in $\RF(\La)$ allow us to compute the lengths $|v_i|$ from the formulae of Definition~\ref{dfn:forms3d}, for example $|v_0|=\sqrt{r_{01}^2+r_{02}^2+r_{03}^2}$.
Up to orientation-preserving isometry, one can fix $v_0$ along the positive $x$-axis in $\R^3$.
The angle $\angle(v_i,v_j)=\arccos\dfrac{v_i\cdot v_j}{|v_i|\cdot|v_j|}\in[0,\pi)$ between the vectors $v_i,v_j$ can be found from the vonorms $v_i^2,v_j^2$ and root product $r_{ij}=\sqrt{-v_i\cdot v_j}$.
A known length $|v_1|$ and angle $\angle(v_0,v_1)$ allow us to fix $v_1$ in the $xy$-plane of $\R^3$.
The vector $v_2$ with a known length $|v_2|$ and two angles $\angle(v_0,v_2)$ and $\angle(v_1,v_2)$ has two symmetric positions with respect to the $xy$-plane spanned by $v_0,v_1$.
These positions can be distinguished by an order of root products in $\RF^+(\La)$ if we reconstruct up to orientaton-preserving isometry. 
The resulting superbases is unique up to isometry by Lemma~\ref{lem:isometric_superbases}.
\end{proof}

\begin{thm}[isometry classification: 3D lattices $\lra$ root forms]
\label{thm:classification3d}
Lattices $\La,\La'\subset\R^2$ are isometric if and only if their root forms coincide: $\RF(\La)=\RF(\La')$ or, equivalently, their coforms and voforms are isomorphic: $\CF(\La)\sim\CF(\La')$, $\VF(\La)\sim\VF(\La')$.
The existence of orientation-preserving isometry is equivalent to $\RF^+(\La)=\RF^+(\La')$.
\bs 
\end{thm}
\begin{proof}
The part \emph{only if} ($\Rightarrow$) means that any isometric lattices $\La,\La'$ have $\RF(\La)=\RF(\La')$.
Lemma~\ref{lem:forms3d_invariants} implies that the root form $\RF(B)$ of an obtuse superbase $B$ is invariant under isometry.  
Theorem~\ref{lem:isometric_superbases} implies $\RF(\La)$ is independent of $B$.
\medskip

The part \emph{if} ($\Leftarrow$) follows from Lemma~\ref{lem:superbase_reconstruction} by reconstructing a superbase of $\La$. 
\end{proof}

Similarly to the 2-dimensional case in \citeasnoun[Definition~7.3]{bright2021easily}, one can visualise root forms of many 3D lattices by projecting two triples $(r_{23},r_{13},r_{12})$ and $(r_{01},r_{02},r_{03})$ from the positive octant to a triangle.
Due to the order $r_{23}\leq r_{13}\leq r_{12}$, after scaling by $(r_{23}+r_{13}+r_{12})^{-1}$ the top triple maps to a point in the quotient triangle $\QT$ with coordinates $x=(\bar r_{12}-\bar r_{13})/2\in[0,\frac{1}{2}]$ and $y=\bar r_{23}\in[0,\frac{1}{3}]$. 
The bottom triple $(r_{01},r_{02},r_{03})$ is not ordered and maps under scaling by $(r_{01}+r_{02}+r_{03})^{-1}$ to a point in the full triangle $\FT$ with coordinates $x=(\bar r_{03}-\bar r_{02})/2\in[-\frac{1}{2},\frac{1}{2}]$ and $y=\bar r_{01}\in[0,1]$.

\section{Easily computable continuous metrics on root forms in dimension 3}
\label{sec:metrics3D}

Any isomorphism on coforms from Definition~\ref{dfn:forms3d} similarly acts on a root form $\RF(\La)$ rearranging root products $r_{ij}$ by one of 24 permutations from the group $S_4$ (for any isometries) or 12 even permutations from the group $A_4$ (for orientation-preserving isometries). 
The root metric is obtained from any distance $d$ between root forms considered as 6-dimensional vectors by minimising over all such permutations.

\begin{dfn}[space $\RFL^{(3)}$ with root metrics $\RM_d(\La,\La')$]
\label{dfn:metrics3d}
For any metric $d$ on $\R^6$, the \emph{root metric} is 
$\RM_d(\La,\La')=\min\limits_{\si\in S_4} d(\RF(\La),\si(\RF(\La'))$, where a permutation $\si$ applies to $\RF(\La')$ as a vector in $\R^6$.
The \emph{orientation-preserving} root metric $\RM_d^+(\La,\La')=\min\limits_{\si\in A_4}d(\RF(\La),\si(\RF(\La')))$ is minimised over even permutations.  
\medskip

If we use the Minkowski $L_q$-norm $||v||_q=(\sum\limits_{i=1}^n |x_i|^q)^{1/q}$ of a vector $v=(x_1,\dots,x_n)\in\R^n$  for any real parameter $q\in[1,+\infty]$, the root metric is denoted by $\CM_q(\La,\La')$.
The limit case $q=+\infty$ means that
$||v||_{+\infty}=\max\limits_{i=1,\dots,n}|x_i|$.
Let $\RFL^{(3)}$ denote the space of \emph{Root Forms of 3-dimensional Lattices} $\La\subset\R^3$, where we can use any of the above metrics satisfying all necessary axioms by Lemma~\ref{lem:metric_axioms}.
\bs
\end{dfn}

The proof of Lemma~\ref{lem:metric_axioms} is almost identical to \citeasnoun[Lemma~8.3]{bright2021easily}.

\begin{lem}[metric axioms for $\CM_d$]
\label{lem:metric_axioms}
For any metric $d$ on $\R^6$, the root metrics $\RM_d$, $\RM_d^+$ from Definition~\ref{dfn:metrics3d} satisfy the metric axioms in Problem~\ref{pro:metric}c.
\bs
\end{lem}

\begin{lem}[Lemma~8.4 in \cite{bright2021easily}]
\label{lem:continuity_products}
Let vectors $u_1,u_2,v_1,v_2\in\R^n$ have a maximum Euclidean length $l$, scalar products $u_1\cdot u_2,v_1\cdot v_2\leq 0$ and be $\de$-close in terms of Euclidean distance: 
$|u_i-v_i|\leq\de$, $i=1,2$.
Then $|\sqrt{-u_1\cdot u_2}-\sqrt{-v_1\cdot v_2}|\leq\sqrt{2l\de}$.
\bs
\end{lem}

Theorems~\ref{thm:superbases->root_forms} and~\ref{thm:root_forms->superbases} show that the 1-1 map $\OSI\lra\LISP\lra\RFL$ established by Theorems~\ref{thm:superbases/isometry} and \ref{thm:classification3d} is continuous in both directions. 

\begin{thm}[continuity of $\OSI\to\RFL$]
\label{thm:superbases->root_forms}
Let lattices $\La,\La'\subset\R^3$ have obtuse superbases $B=\{v_i\}_{i=0}^3$, $B'=\{u_i\}_{i=0}^3$ whose vectors have a maximum length $l$ and $|u_i-v_i|\leq\de$ for some $\de>0$, $i=0,1,2$.
Then $\RM_q(\RF(\La),\RF(\La'))\leq 6^{1/q}\sqrt{2l\de}$ for any $q\in[1,+\infty]$, where $6^{1/q}$ is interpreted for $q=+\infty$ as $\lim\limits_{q\to+\infty}6^{1/q}=1$. 
The same upper bound holds for the orientation-preserving metric $\RM_q^+$.
\bs
\end{thm}
\begin{proof}
Lemma~\ref{lem:continuity_products} implies that the root products $r_{ij}=\sqrt{-v_i\cdot v_j}$ and $\sqrt{-u_i\cdot u_j}$ of the superbases $B,B'$ differ by at most $2l\de$ for any pair $(i,j)$ of indices.
Then the $L_q$-norm of the vector difference in $\R^3$ is
$\RM_q(\RF(\La),\RF(\La'))\leq 6^{1/q}\sqrt{2l\de}$ for any $q\in[1,+\infty]$.
By Definition~\ref{dfn:metrics3d}, the root metric $\RM_q$ is minimised over permutations of $S_4$ (or $A_4$ for the orientation-preserving metric $\RM_q^+$), so the upper bound still holds.
\end{proof}

Theorem~\ref{thm:superbases->root_forms} is proved for the $L_q$ norm only to give the explicit upper bound for $\RM_q$.
A similar argument proves continuity for $\RM_d$ with any metric $d$ on $\R^3$ satisfying $d(u,v)\to 0$ when $u\to v$ coordinate-wise. 
Theorem~\ref{thm:root_forms->superbases} is stated for $L_{+\infty}$ only for simplicity, because all Minkowski norms in $\R^n$ are topologically equivalent due to $||v||_q\leq ||v||_{r}\leq n^{\frac{1}{q}-\frac{1}{r}}||v||_q$ for any $1\leq q\leq r$ \cite{norms}. 

\begin{thm}[continuity of $\RFL\to\OSI$]
\label{thm:root_forms->superbases}
Let lattices $\La,\La'\subset\R^3$ have $\de$-close root forms, so $\RM_{\infty}(\RF(\La),\RF(\La'))\leq\de$.
Then $\La,\La'$ have obtuse superbases $B$, $B'$ that are close in the $L_{\infty}$ metric on the space $\OSI$ so that $L_{\infty}(B,B')\to 0$ as $\de\to 0$.
The same conclusion holds for the orientation-preserving metrics $\RM_{\infty}^+$ and $L_{\infty}^+$.
\bs
\end{thm}
\begin{proof}
Superbases $B=\{v_i\}_{i=0}^3$, $B'=\{u_i\}_{i=0}^3$ can be reconstructed from the root forms $\RF(\La),\RF(\La')$ by Lemma~\ref{lem:superbase_reconstruction}.
By applying a suitable isometry of $\R^3$, one can assume that $\La,\La'$ share the origin and the first vectors $v_0,u_0$ lie in the positive $x$-axis.
Let $r_{ij},s_{ij}$ be the root products of $B,B'$ respectively.
Definition~\ref{dfn:forms3d} implies that $v_i^2=r_{ij}^2+r_{ik}^2+r_{il}^2$ and $u_i^2=s_{ij}^2+s_{ij}^2$ for distinct indices $i,j,k,l\in\{0,1,2,3\}$, for example if $i=0$ then $j=1$, $k=2$, $l=3$. 
For any continuous transformation from $\RF(\La)$ to $\RF(\La')$, all root products have a finite upper bound $M$, which is used below:
$$|v_i^2-u_i^2|=|(r_{ij}^2+r_{ik}^2)-(s_{ij}^2+s_{ik}^2)|\leq
|r_{ij}^2-s_{ij}^2|+|r_{ik}^2-s_{ik}^2|\leq $$
$$(r_{ij}+s_{ij})|r_{ij}-s_{ij}|+(r_{ik}+s_{ik})|r_{ik}-s_{ik}|\leq 
(r_{ij}+s_{ij})\de+(r_{ik}+s_{ik})\de\leq 4M\de.$$

Since at least two continuously changing conorms should be strictly positive to guarantee positive lengths of basis vectors by Definition~\ref{dfn:forms3d}, there is a minimum length $a>0$ of all basis vectors during a transformation $\La'\to\La$.
Then $||v_i|-|u_i||\leq\dfrac{4M\de}{|v_i|+|u_i|}\leq\dfrac{2M}{a}\de$.
Since the first basis vectors $v_0,u_0$ lie in the positive horizontal axis, the lengths can be replaced by vectors: $|v_0-u_0|\leq\dfrac{2M}{a}\de$, so $|v_0-u_0|\to 0$ as $\de\to 0$.
\medskip

Up to orientation-preserving isometry and keeping both $v_0,u_0$ fixed in the positive $x$-axis, one can put the vectors $v_1,u_1$ into the $xy$-plane of $\R^3$. 
Then $v_1,u_1$ can have a non-zero angle equal to the difference $\al_1-\be_1$ of the angles from the positive $x$-axis in $\R^3$ to $v_1,u_1$, respectively.
These angles are expressed via the root products as follows:
$$\al_i=\arccos\dfrac{v_0\cdot v_i}{|v_0|\cdot|v_i|}
=\arccos\dfrac{-r_{0i}^2}{\sqrt{r_{01}^2+r_{02}^2}\sqrt{r_{ij}^2+r_{ik}^2}},\leqno{(\ref{thm:root_forms->superbases}a)}$$
$$
\be_i=\arccos\dfrac{u_0\cdot u_i}{|u_0|\cdot|u_i|}
=\arccos\dfrac{-s_{0i}^2}{\sqrt{s_{01}^2+s_{02}^2}\sqrt{s_{ij}^2+s_{ik}^2}}\leqno{(\ref{thm:root_forms->superbases}b)}
$$ 
for distinct $i,j,k\in\{1,2,3\}$.
If $\de\to 0$, then $s_{ij}\to r_{ij}$ and $\al_i-\be_i\to 0$ for all indices, because all the above functions are continuous for $|u_j|,|v_j|\geq a$, $j=0,1,2,3$.
\medskip

Then we estimate the squared length of the difference by using the scalar product:
$$|v_i-u_i|^2=v_i^2+u_i^2-2u_iv_i
=(|v_i|^2-2|u_i|\cdot |v_i|+|u_i|^2)+2|u_i|\cdot |v_i|-2|u_i|\cdot |v_i|\cos(\al_i-\be_i)=$$
$$
=(|v_i|-|u_i|)^2+2|u_i|\cdot |v_i|(1-\cos(\al_i-\be_i))
=(|v_i|-|u_i|)^2+|u_i|\cdot |v_i|4\sin^2\dfrac{\al_i-\be_i}{2}
\leq$$
$$\leq (|v_i|-|u_i|)^2+|u_i|\cdot |v_i|4\left(\dfrac{\al_i-\be_i}{2}\right)^2
=(|v_i|-|u_i|)^2+|u_i|\cdot |v_i|(\al_i-\be_i)^2,$$
where we have used that $|\sin x|\leq|x|$ for any real $x$. 
The upper bound $M$ of all root products guarantees a fixed upper bound for lengths $|u_i|,|v_i|$.
The above arguments starting from formulae~(\ref{thm:root_forms->superbases}a,b) hold for any index $i=1,2,3$.
For $i=1$, if $\de\to 0$ then $|v_1|-|u_1|\to 0$ and $\al_1-\be_1\to 0$ as proved above, so we conclude that $u_1\to v_1$.
\medskip

The vectors $u_2,v_2$ may not be in the same $xy$-plane in $\R^3$.
If they are in the same line, the difference $u_2-v_2$ tends to 0 as $|u_2|-|v_2|\to 0$.
Otherwise $v_2,u_2$ span a plane $P_2$ intersecting the $xy$-plane ijn a line $L_2$.
A unit length vector $w_2$ along $L_2$ can be expressed as linear combinations $a_0v_0+a_1v_1=b_0v_0+b_1v_1$ for some coefficients $a_0,a_1,b_0,b_1\in\R$.
The above arguments now work for angles measured from $w_2$ (instead of $u_0$ and $v_0$) to $v_2,u_2$.
Since we already know that $u_0\to v_0$ and $u_1\to v_1$ as $\de\to 0$, we get $a_0\to b_0$ and $a_1\to b_1$.
All arguments about continuity of scalar products and angles work when the vectors $u_0,v_0$ in the $x$-axis are replaced by $w_2$ in the $xy$-plane.
\medskip

So we get $u_2\to v_2$, similarly $u_3\to v_3$, and finally $L_{\infty}(B,B')\to 0$ as $\de\to 0$.
\end{proof}

\section{Visualisation of large families of lattices from the CSD and conclusions}
\label{sec:conclusions}

\begin{figure}
\label{fig:CSDorthorhombic3D}
\caption{
Density plot of all 145,199 primitive orthorhombic lattices in the CSD.
Any such lattice is represented by a triple of side lengths $a=r_{01}\leq b=r_{02}\leq b=r_{03}$, which under scaling by $(a+b+c)^{-1}$ are projected to the quotient triangle $\QT$ .}
\includegraphics[width=1.0\textwidth]{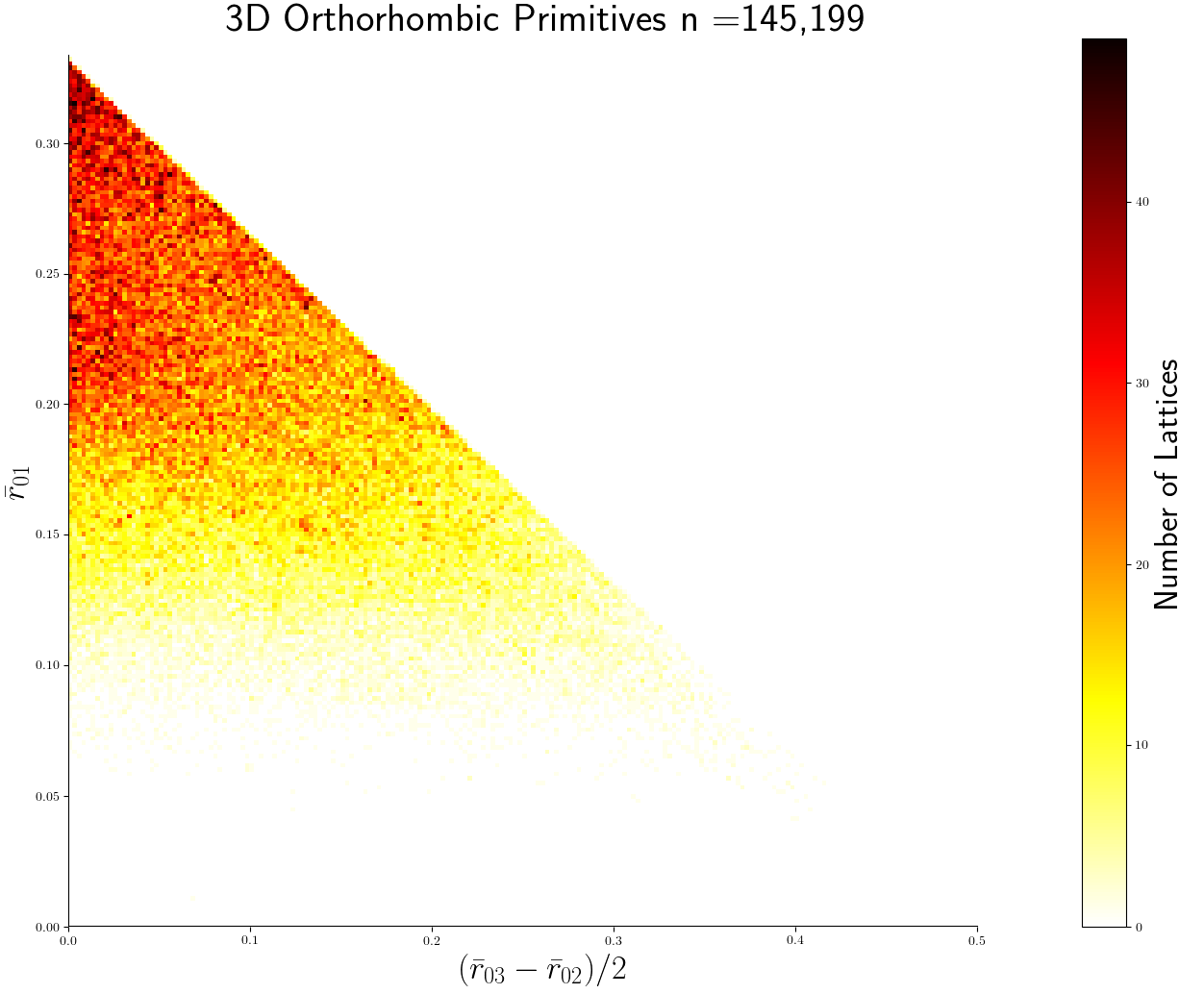}
\end{figure}

The Cambridge Structural Database (CSD) has about 145K crystals whose lattices are primitive orthorhombic. 
To represent such a large number of real lattices, we subdivide the quotient triangle  into a $200 \times 200$ grid and count lattices parameters fall into each pixel.
These counts (from 0 to 75) are represented the colour bar on the right hand side of Fig.~\ref{fig:CSDorthorhombic3D}. 
The resulting plot shows high density pixels close to the top vertex, which represents cubical lattices.
The white region for $\bar r_{01}<0.1$ indicates that there are much fewer orthorhombic lattices with one side considerably shorter than others.

For more generic triclinic lattices $\La$, the root form $\RF(\La)$ consists of two rows: the top ordered triple $r_{23}\leq r_{13}\leq r_{12}$ and the bottom unordered triple $(r_{01},r_{02},r_{03})$.

\begin{figure}
\label{fig:CSDtriclinic3D}
\caption{Scatter plot for all triclinic lattices $\La$ from the CSD.
\textbf{Top}: the ordered top rows of root forms $\RF(\La)$ are projected to the quotient triangle $\QT$.
\textbf{Bottom}: the unordered bottom rows of root forms $\RF(\La)$ are projected to the full triangle $\FT$.
}
\includegraphics[width=0.9\textwidth]{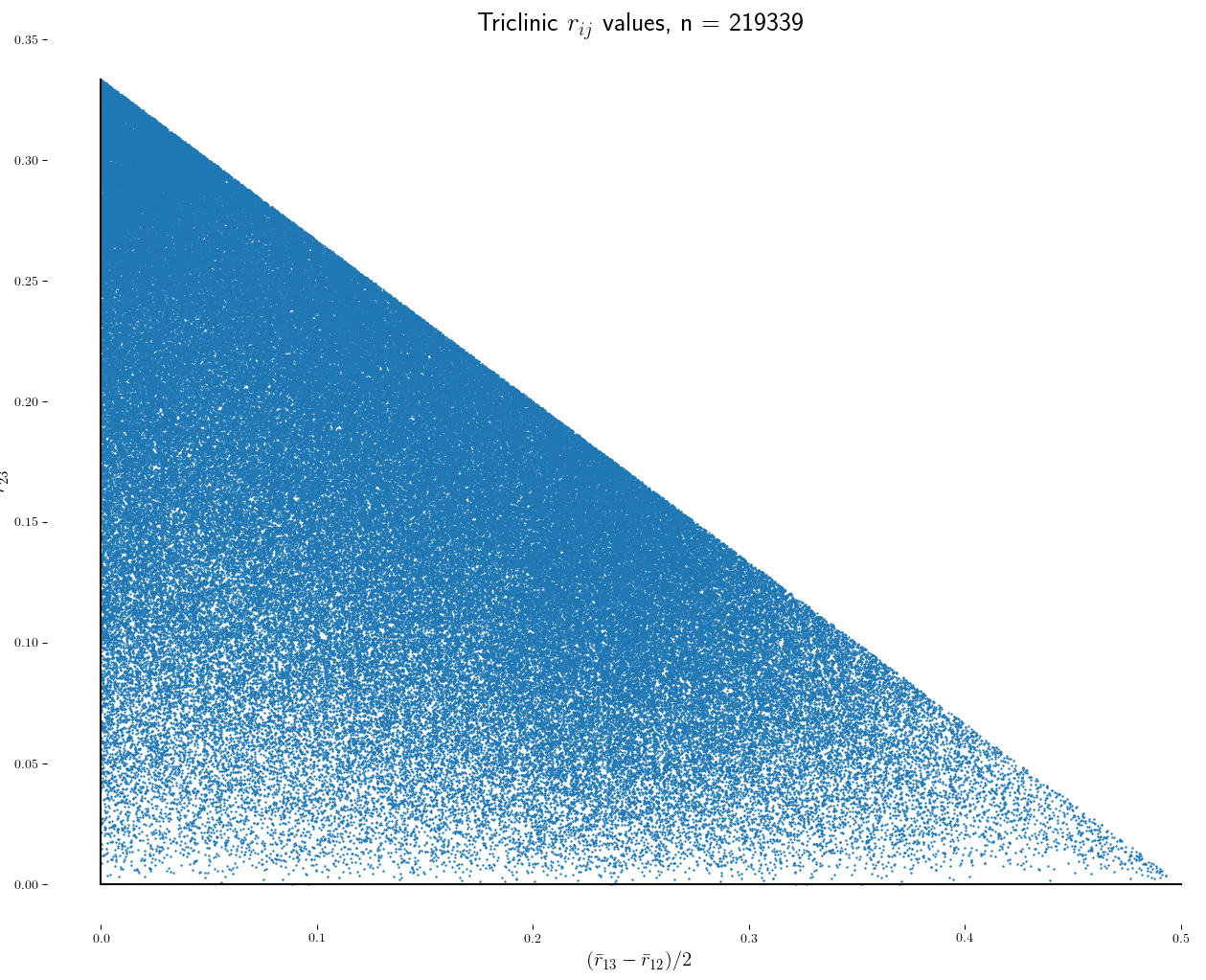}

\includegraphics[width=\textwidth]{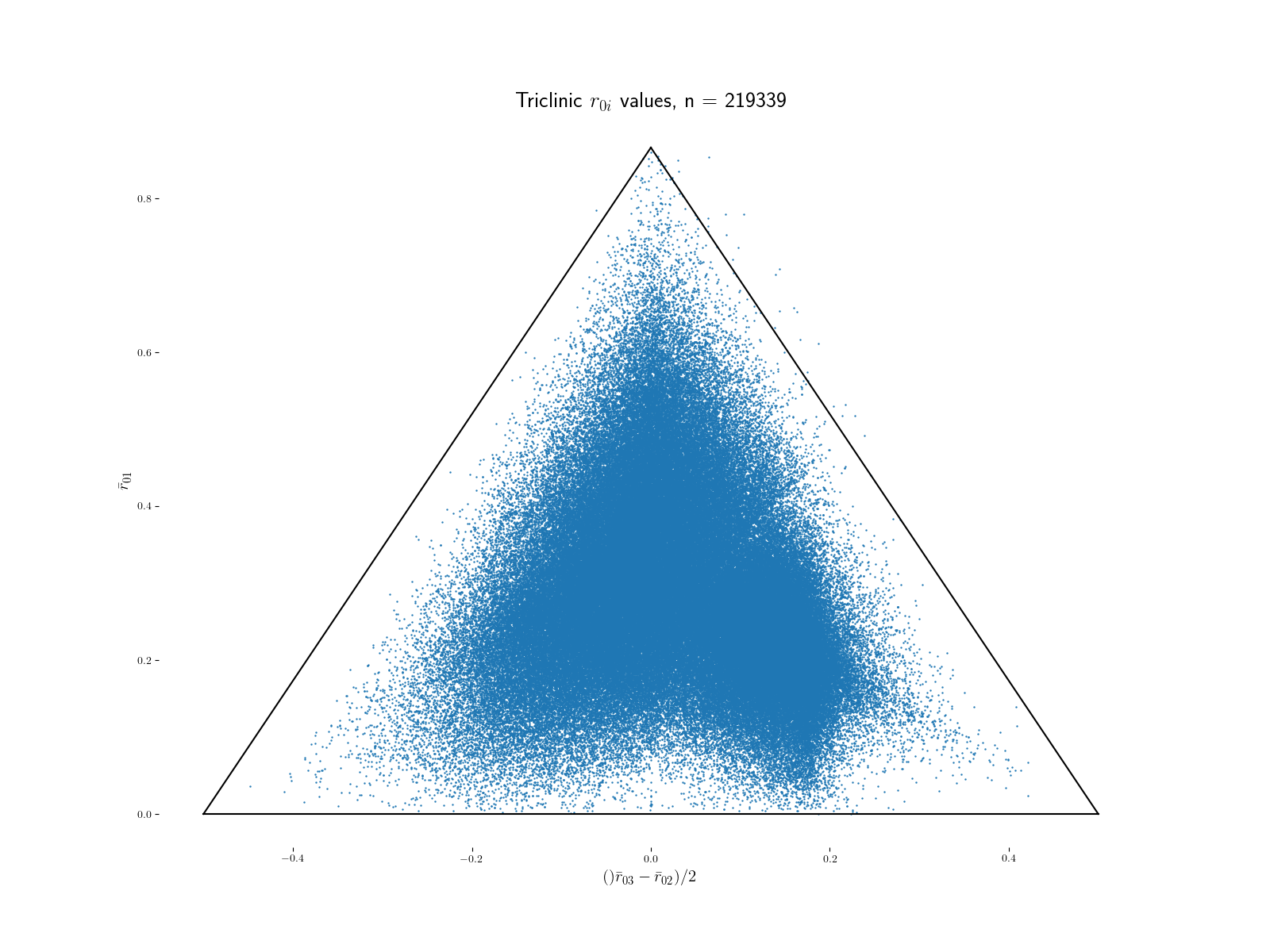}
\end{figure}

The large-scale visualisations confirm that real lattices form a continuum, which further motivates a continuous crystallography, see \citeasnoun[section~10]{bright2021easily}.

\appendix



\ack{Acknowledgements}.
We are grateful to many colleagues for helpful discussions during the MACSMIN 2021 conference (Mathematics and Computer Science for Materials Innovation, http://kurlin.org/macsmin.php\#2021), especially to Larry Andrews and Herbert Bernstein.
The research has been supported by the £3.5M EPSRC grant ``Application-driven Topological Data Analysis'' (2018-2023, EP/R018472/1).



\referencelist

\end{document}